\documentstyle[12pt,aasms4]{article}
\def\msun{{\,M_\odot}}
\def\simlt{\lower.5ex\hbox{$\; \buildrel < \over \sim \;$}}
\def\simgt{\lower.5ex\hbox{$\; \buildrel > \over \sim \;$}}

\def\pc{{\rm\,pc}}

\def\yr{{\rm\,yr}}

\def\kms{{\rm\,km\,s^{-1}}}
\def\mdot{{\rm\,\msun\,yr^{-1}}}
\def\gms{{\rm\,g\,s^{-1}}}
\def\gcm3{{\rm\,g\,cm^{-3}}}
\def\ncm3{{\rm\,cm^{-3}}}
\def\kelvin{{\rm\,K}}

\def\>{$>$}
\def\<{$<$}

\def\refbook#1{\refindent#1}
\def\refindent{\par\noindent\hangindent=3pc\hangafter=1 }
\def\aa#1#2#3{\refindent#1, A\&A, {\bf#2}, #3.}
\def\aalett#1#2#3{\refindent#1, A\&A {\it (Letters)}, {\bf#2}, #3.}

\def\apj#1#2#3{\refindent#1, {\it ApJ}, {\bf#2}, #3.}
\def\apjlett#1#2#3{\refindent#1, {\it ApJ (Letters)}, {\bf #2}, #3.}

\def\baas#1#2#3{\refindent#1, BAAS, #2, #3}

\def\mnras#1#2#3{\refindent#1, {\it MNRAS}, {\bf#2}, #3.}
\def\nature#1#2#3{\refindent#1, {\it Nature}, {\bf #2}, #3.}

\lefthead{Coker \& Melia}
\righthead{Bondi-Hoyle Accretion onto Sgr A*}

\begin{document}

\title{Hydrodynamical Accretion Onto Sgr A* From\\
       Distributed Point Sources}

\author{Robert F. Coker\altaffilmark{1}$^*$ and Fulvio Melia\altaffilmark{2}$^{*\dag}$}
\affil{$^*$Physics Department, The University of Arizona, Tucson, AZ 85721}
\affil{$^{\dag}$Steward Observatory, The University of Arizona, Tucson, AZ 85721}

\altaffiltext{1}{NASA GSRP Fellow.}
\altaffiltext{2}{Presidential Young Investigator.}

\begin{abstract}
Spectral and kinematic studies suggest that the nonthermal radio source
Sgr A*, located at the center of the Milky Way, is a supermassive compact
object with a mass $\sim 2-3\times{10}^6\msun$.  Winds from nearby stars,
located $\approx 0.06$ pc to the east of Sgr A*, should, in the absence of
any outflow from the putative black hole itself, be accreting onto this object.
We report the results of the first 3D Bondi-Hoyle hydrodynamical numerical
simulations of this process under the assumption that the Galactic center
wind is generated by several different point sources (here assumed to
be 10 pseudo-randomly placed stars).  Our results
show that the accretion rate onto the central object can be higher than in the
case of a uniform flow since wind-wind shocks dissipate some of the bulk
kinetic energy and lead to a higher capture rate for the gas.  However,
even for this highly non-uniform medium, most of the accreting gas carries
with it a relatively low level of specific angular momentum, though
large transient fluctuations can occur.  Additionally, the post-bow-shock
focusing of the gas can be substantially different than that for a uniform flow,
but it depends strongly on the stellar spatial distribution.
We discuss how this affects the morphology of the gas in the inner 0.15 pc of
the Galaxy and the consequences for accretion disk models of Sgr A*.
\end{abstract}


\section{Introduction}
Sgr A* may be a massive ($\sim 2-3\times 10^6\;M_\odot$) point-like
object dominating the gravitational potential in the inner $\la 0.5$ pc
region of the Galaxy.  This inference is based on the
large proper motion of nearby stars (\cite{HA95}; \cite{EG97};
\cite{GEOE97}),
the spectrum of Sgr A* (\cite{MJN92}), its low proper motion ($\simlt 20 \kms$;
\cite{B96}), and its unique location (\cite{LAS91}).  The gaseous
flows in this region are themselves rather complex, and key
constituents appear to be the cluster of mass-losing, blue, luminous
stars comprising the IRS 16 assemblage, which is located within several
arc seconds ($1^{\prime\prime} \approx$ 0.04 pc in projection at the distance
to the Galactic center) from Sgr A*.  Measurements of high outflow velocities
associated with IR sources in Sgr A West (\cite{K91}) and in IRS 16 (\cite{G91}),
the $H_2$ emission in the circumnuclear disk (CND) from molecular gas being
shocked by a nuclear mass outflow (\cite{G86}), broad Br$\alpha$,
Br$\gamma$ and He I emission lines 
(\cite{HKS82}; \cite{AHH90}; \cite{G91}), and radio continuum observations of IRS 7
(\cite{YM91}), provide clear evidence of a hypersonic wind, with velocity
$v_w \sim500-1000\; \kms$, a number density $n_w\sim10^{3-4}\;\ncm3$, and a total
mass loss rate $\dot M_w\sim3-4\times10^{-3}\mdot$, pervading the inner
parsec of the Galaxy.  Many of Sgr A*'s radiative characteristics may be
due to its accretion of the IRS 16 wind.

In the classical Bondi-Hoyle (BH) scenario (\cite{BH44}), the mass accretion
rate for a uniform hypersonic flow is $\dot M_{BH} = \pi {R_A}^2 m_H n_w v_w$,
in terms of the accretion radius $R_A \equiv 2 G M / {v_w}^2$.
At the Galactic center, with $n_w \sim 5.5 \times 10^3 \ncm3$
and $v_w \sim700 \kms$, we would therefore expect an accretion rate
$\dot M_{BH} \sim 10^{22} \gms$ onto the black hole, with a capture
radius $R_A \sim .02 \pc$.  Since this accretion rate is sub-Eddington
for a one million solar mass object, the accreting gas is mostly
unimpeded by the escaping radiation field and is thus
essentially in hydrodynamic free-fall starting at $R_A$.
Our initial numerical simulations of this process, assuming a highly
simplistic uniform flow past a point mass (\cite{RM94}; \cite{CM96}) have
verified these expectations.

On the other hand, the nature of Bondi-Hoyle accretion onto a
point like object also presents somewhat of a challenge in understanding
what happens to the gas as it settles down into a planar configuration
close to the event horizon.  Fluctuations beyond the 
bow shock (located at $\sim R_A$) produce
a transient accretion of net angular momentum that ought to result in
the formation of a temporary (albeit small) disk. The circularization
radius is $r_c \approx 2 \lambda^2 r_g$, where $r_g = 2 G M / c^2$ is
the Schwarzschild radius and $\lambda$ is the specific angular momentum
in units of $c r_g$.  Our simulations of the BH accretion from a uniform
flow suggest that in this configuration $\langle\lambda\rangle\sim 3-20$.
More realistically, the inflow itself carries angular momentum, so that
the formation of a disk-like structure at small radii (i.e., $r\approx
10^{2-3}\;r_g$) may be difficult to avoid (\cite{M94}).  However,
the observations do not appear to favor the presence of a {\it standard}
$\alpha$-disk at small radii (\cite{M96}).  The
current upper limits on the infrared flux from Sgr A* (\cite{M97}) suggest
that either (1) the circularized flow does not form an $\alpha$-disk,
but rather advects most of its dissipated energy through the event
horizon (e.g., \cite{NYM95}), (2) the Bondi-Hoyle flow merges into
a massive, fossilized disk, storing most of the deposited matter
at large radii (\cite{FM97}), or (3) Sgr A* is not a point-like object
(\cite{HA95}).

Added to this is the fact that in reality the flow past Sgr A* is not
likely to be uniform.  For example, one might expect many shocks
to form as a result of wind-wind collisions within the IRS 16 comples, 
even before the plasma reaches
$R_A$.  With this consequent loss of bulk kinetic energy, it would not
be surprising to see the black hole accrete at an even larger rate than in
the uniform case.  The implications for the spectral characteristics of
Sgr A*, and thus its nature, are significant.  We have therefore
undertaken the task of simulating the BH accretion from the spherical winds
of a distribution of 10 individual point sources located at an average
distance of a few $R_A$ from the central object. As we shall
see below, the accretion rate depends not only on the distance
of the mass-losing star cluster from the accretor but also on the relative
spatial distribution of the sources.  As suggested by
related work involving linear gradients (\cite{RA95}), the average value
of $\lambda$ is also larger than that of the uniform case,
exhibiting large temporal fluctuations, but still not as large as one might expect.
In this {\sl Letter} we present the results of
the first 3D hydrodynamical calculations of multiple-source stellar winds
accreted by a point mass.

\section{The Calculational Algorithm and Physical Setup}
Although we here attempt to model the wind source more realistically
than in previous work, we shall still focus on the hydrodynamical aspects of the
flow and thus we do not include heating or cooling (see, e.g., \cite{M94}).
Nor do we include any relativistic effects.  Instead,
we assume the medium to be an unmagnetized, polytropic gas with pressure
$P = (\gamma - 1)\rho e$ where $\rho$, and $e$ are the mass and
internal energy densities, respectively, and $\gamma = 4/3$ is the adiabatic
index.  This value of $\gamma$ was chosen so that the sonic point
$d_s = R_A \times (5-3\gamma)/4$ occurs outside the accretor, which
typically is modeled with a radius $\sim 0.1R_A$.  Some earlier work (e.g.,
\cite{P89}) found that varying $\gamma$ affects the flow pattern
significantly, particularly at radii much smaller than $R_A$,
while others (e.g., \cite{R96}) found that it does not, particularly
at the larger radii considered here.

We use the numerical algorithm ZEUS-3D, a general purpose code for
MHD fluids developed at NCSA (\cite{No94}).  
The calculations are carried out within a cubical domain of solution
with ${112}^3$ active zones geometrically scaled so that the central zones
are 1/32 times the size of the outermost zones, closely
mimicking the ``multiply nested grids'' arrangement used by
other researchers (e.g., \cite{RM94}).  This allows for maximal
resolution of the accretor within the computer memory limits available while
sufficiently resolving the wind sources and minimizing zone-to-zone boundary
effects. The total volume is $(16 R_A)^3$ or $\sim(0.28 \pc)^3$ using a point mass
$M = 1\times{10}^6\msun$ located at the origin and a Mach 10 Galactic center wind with
$v_w = 700\;\kms$.  

The initial density is set to a small
value and the velocity is set to zero.  The internal energy density
is chosen such that the temperature is $\sim10^4\;\kelvin$.
Free outflow conditions are imposed on the outermost zones and each
time step is determined by the Courant condition with a Courant number of 0.5.
The 10 identical stellar wind sources are modeled by forcing
the velocity in 10 subregions of $5^3$ zones to
be constant (at $v_w$) while the densities
in these subvolumes are set so that the total mass flow into
the volume of solution, $\dot M_w$, is 3$\times10^{-3}\mdot$.  
Recent observations
(\cite{N94}) suggest that the IRS 16 wind is somewhat colder than the
temperature assumed here, with a Mach number of $\sim$ 30, and that it
may originate from more than 15 sources of varying strength rather
than 10 sources of equal strength.  This level of detail will
be addressed in future work.

To gauge the dependence of our results on the source configuration,
we here simulate the BH process using two different stellar distributions.
For numerical reasons, the sources need to be placed a minimum of 8 zones
($\simlt .2 R_A$) apart so no attempt is made here to imitate exactly
the distribution within the IRS 16 cluster.  The {\it average} location for the
sources in run 1 is $\sim 4 R_A \hat{z}$ (with the accretor located at the origin)
and $\sim 3 R_A \hat{z}$ for run 2.
For run 1, the sources are distributed fairly randomly with x, y, and z being
allowed to vary by $\pm3 R_A$ of the average location.  For run 2,
x and y are allowed to be as much as 5 $R_A$ from the average location
while z can only vary by $\pm1 R_A$.  These distributions are chosen
to represent the extremes of a spherical stellar distribution versus a
more or less planar one.  For comparison, IRS 16 NW and IRS 16 NE, two bright
members of the IRS 16 cluster, are $\sim3$ and $\sim7 R_A$ away
(in projection) from Sgr A* (\cite{M97}).  Each simulation is evolved
for 2000 years.  The wind
crossing time is $\sim$ 400 years and equilibrium,
the point at which the original gas is swept
out of the volume of solution and the mass accretion rate stabilizes,
appears to be reached within
2 crossing times, or $\sim$ 800 years.

\section{Comparison of the Flow Patterns}
Figure 1 shows a logarithmic negative grey scale image of the density
profile for a slice running through the center of the accretor,
for run 1 taken 2000 years after the winds are ``turned on''.
The image is 6 $R_A$ on a side with the $0.1 R_A$ radius
accretor located at the center.  One stellar source
can be clearly seen to the upper right of the accretor; the other 9 sources are either not
in the plane of the slice or are off to the right.  The image spans 4 magnitudes
of density with white being $\simlt0.01 n_w$ and black being $\simgt10 n_w$.
Figure 2 is a similar image for run 2.  Note the large density
fluctuations across the overall region.  Also, the flow pattern is clearly
different for the two configurations.  
However, once the stellar winds have cleared the
region of the original low density gas, both simulations point to
an overall average density ($\sim10^3\ncm3$) in agreement with observations.

A substantial difference between these point source simulations and previous
uniform calculations is that although there are transient fragmentary shocks
extending from the accretor boundary (see Figure 1), there is no large-scale
bow shock structure.
There are, however, bow shocks around some of the stellar
sources as their wind impacts the stronger accumulated wind of the other sources
(see the source in Figure 1),
and, for run 2, a semi-stable partial bow shock appears in
front of the accretor (see Figure 2).
The wind-wind collisions convert kinetic energy into thermal energy with the result that
a larger fraction of the gas is captured by the central engine. 
Although this effect is likely
to be sensitive to the actual stellar distribution, a larger density of point
sources should produce more wind-wind collisions and perhaps further
raise the accretion rate.

\section{The Accretion of Mass and Angular Momentum onto the Central Object}
For a wind originating from a single point source located at a distance $R$ 
(\> $R_A$) from the accretor, $\dot M$ is less than $\dot M_{BH}$ due to 
the divergent flow.  Specifically, for $R= 4 R_A$, $\dot M \sim 0.15\;
\dot M_{BH}$.  However, wind-wind collisions from multiple sources
reduce the effective $v_w$, thereby increasing
the effective $R_A$ so that all of the shocked gas will
then accrete and more than compensate for the geometric losses.
In Figure 3 we present the mass accretion rate, $\dot M$, and the accreted
specific angular momentum, $\lambda$, versus time for run 1,  starting
2 crossing times ($\sim$ 800 years) after the winds are ``turned on''.
Figure 4 presents the results for run 2.  The
average values for the mass accretion rate
once the systems have reached equilibrium are
$\dot M = 2.1 \pm 0.3 \dot M_{BH}$ for run 1 and $1.1 \pm 0.2 \dot M_{BH}$ for run 2.

The mass accretion rate shows high frequency temporal fluctuations (with a
period of $\simlt 0.25 \yr$) due to the finite numerical
resolution of the simulations.  The low frequency aperiodic
variations (on the order of $20\%$ in amplitude) reflect the time dependent
nature of the flow.  Thus, the mass accretion rate onto the central object,
and consequently the emission arising from within the accretor boundary, is
expected to vary by $\simlt 20-40\%$ (since in some models
$L$ may vary by as much as $\propto {\dot M}^2$) over the corresponding
time scale of $<100$ years, even though the mass flux from the stellar
sources remains constant.

Similarly, for run 1, the accreted $\lambda$ can vary by $50\%$ over 
$\simlt$ 200 years with an average equilibrium value of $37 \pm 10$.
For run 2, $\langle\lambda\rangle = 62 \pm 5$ with smaller amplitude long term variability.
Since previous uniform simulations resulted in $\langle\lambda\rangle \sim 3-20$, it appears
that even with a large amount of angular momentum present in the wind, relatively little
specific angular momentum is accreted.  This is understandable
since clumps of gas with a high specific angular momentum do not penetrate within 1 $R_A$.
The variability in the sign of the components of $\lambda$ suggests
that if an accretion disk forms at all, it dissolves, and reforms (perhaps)
with an opposite sense of spin on a time scale of $\sim 100$ years.

\section{Discussion, Conclusions, and Future Work}
A variety of accretion scenarios for Sgr A* have been proposed over
the years (\cite{M94}; \cite{FO94}; \cite{NYM95}; \cite{Be96}), each with
its own restrictions on $\dot M$ and $\lambda$.  While the accreted
specific angular momentum determined in the present simulations is
an order of magnitude too small to support the fossil disk scenario
(since then the energy liberated as the wind impacts the fossil disk should
be visible; \cite{FM97}), it is still large enough that any standard
$\alpha$-disk would be easily detectable (\cite{M94}).
The advection dominated disk scenario (\cite{NYM95}), while
permitting a large range of values for $\lambda$, requires
an accretion rate of $\simlt 10^{-5}\mdot$ or roughly $0.1 \dot M_{BH}$,
which is 10-20 times smaller than the value derived here.
In addition, our $\dot M$ is more than $10$ times larger than that
permitted by the ``mono-energetic'' electron model of Sgr A* 
(\cite{Be96}).  It appears that additional
work is needed to reconcile disk models with the fact that the observed
multiple wind sources result in a large mass accretion rate onto the central
engine, if its mass is $\sim {10}^6\msun$.  In view of this, it may not
be unreasonable to conjecture that in fact no flattened disk actually
forms in Sgr A*, but rather that the excess angular momentum is dissipated
in a quasi-spherical configuration and that the thermalized energy is then
advected inwards through the event horizon before the gas settles onto a plane
(\cite{Me92}).

These simulations suggest that the $\sim 0.1 \pc$ region of the Galaxy, centered
on the wind sources, is swept
clear of gas, leaving a hot, low density gas filling the central cavity.  This
is consistent with observations of the region within the CND (\cite{YW93}), and
may be acting in concert with other mechanisms to produce the sharp inner
edge of the CND (e.g., the abrupt change in gravitational potential;
\cite{D89}). Additionally, a tongue of hot, dense gas has been observed
that connects members of the IRS 16 cluster to Sgr A* (\cite{G91}).  It is
worthwhile noting that the images in Figures 1 and 2 show ridges of
dense gas connecting the sources in the figures to the accretor.

\section{Acknowledgments}
This work was partially supported by NASA under grants NAGW-2518
and NGT-51637, and utilized the Origin 2000 computer system in Friendly User 
mode at the National Center for Super computing Applications, University of 
Illinois at Urbana-Champaign.  

{}

\clearpage

\figcaption[]{A logarithmic negative grey scale snapshot, 6 $R_A$ on a side,
of the gas density
for run 1 taken 2000 years after the winds are ``turned on''.
It is a slice in the $\hat x-\hat z$ plane (with $\hat z$ to the right)
through the accretor (marked with a filled white circle).
White corresponds to a number density $\simlt 0.01 n_w$ and black
corresponds to a number density $\simgt 10 n_w$.\label{fig1}}

\figcaption[]{Same as Figure 1 but for run 2 at 2000 years.\label{fig2}}

\figcaption[]{Accretion results for run 1.
The upper solid curve is accreted specific angular momentum $\lambda$ (in units
of $cr_g$).  The scale for $\lambda$ is on the left side.  The lower dotted curve
is the mass accretion rate $\dot M$ (${10}^{-4}\mdot$) versus time.
The scale for $\dot M$ is shown on
the right side.\label{fig3}}

\figcaption[]{Accretion results for run 2.\label{fig4}}

\clearpage

\plotone{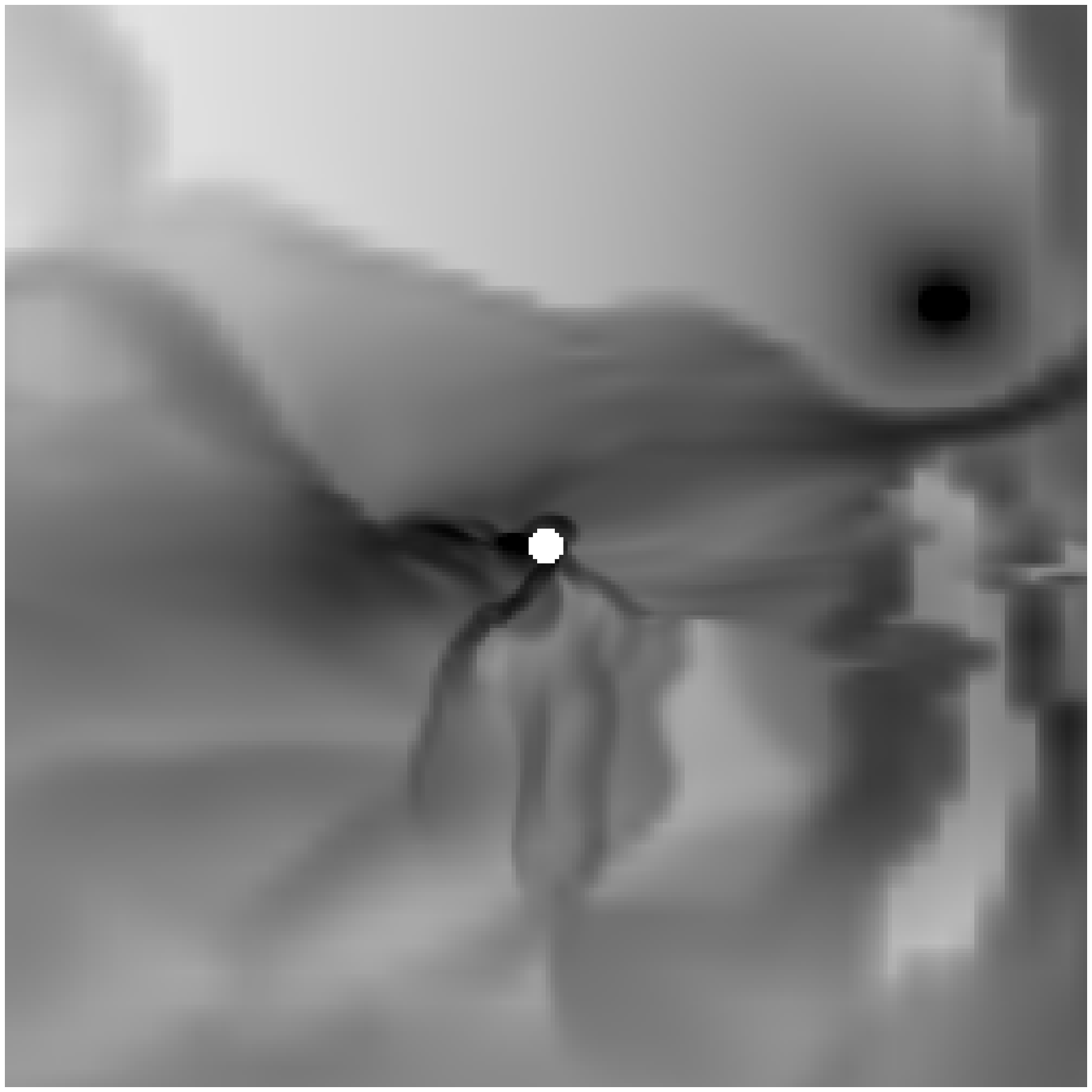}

\clearpage

\plotone{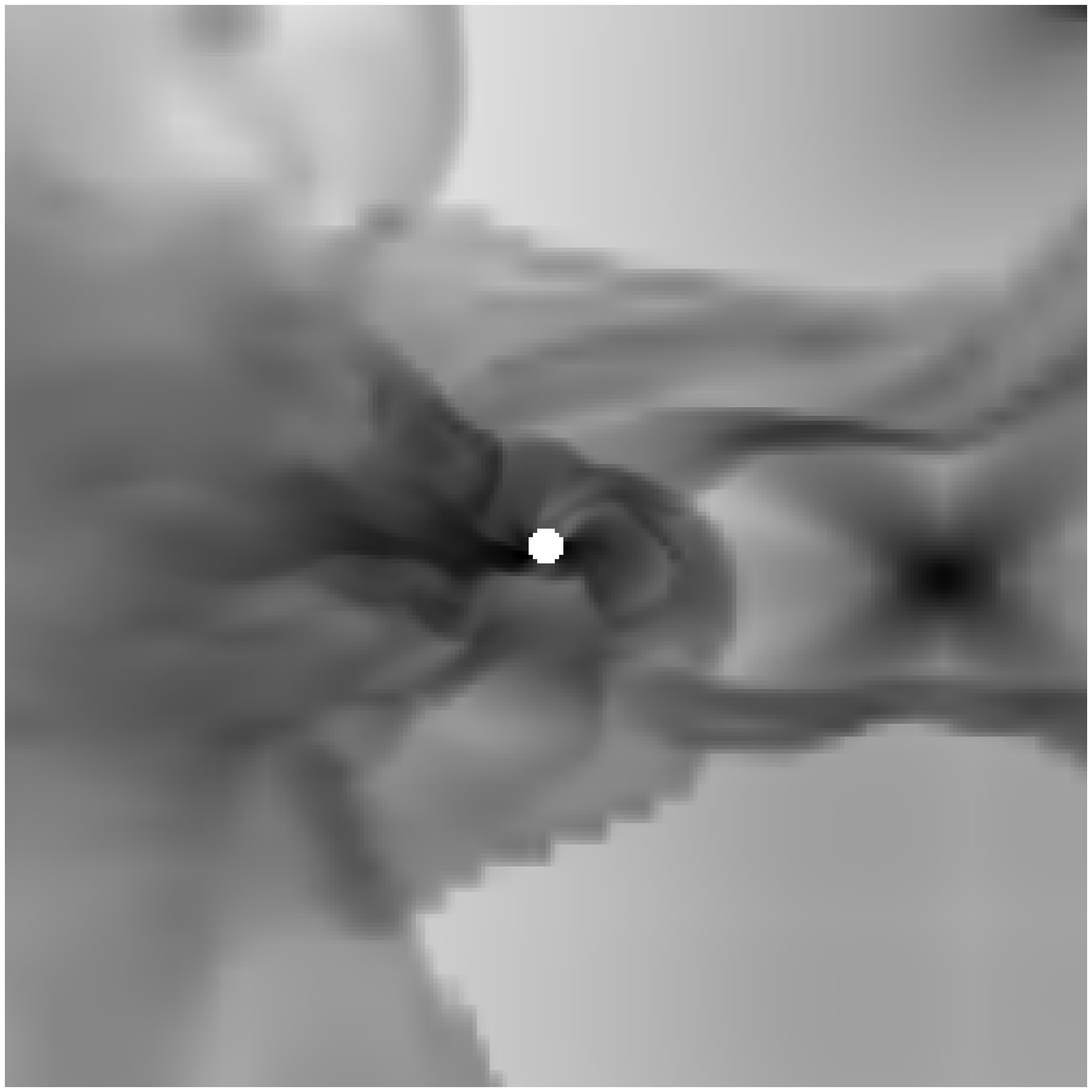}

\clearpage

\plotone{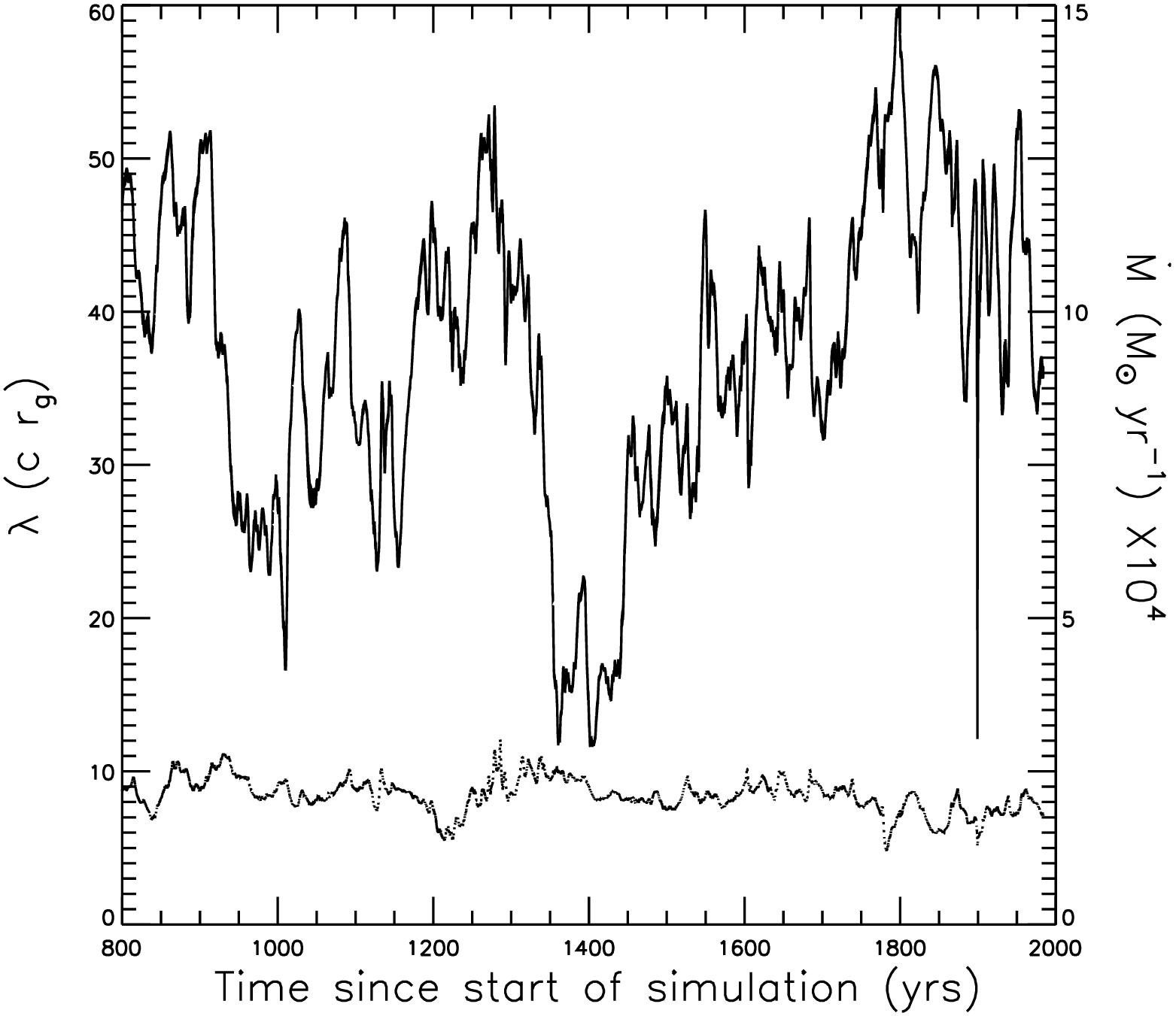}

\clearpage

\plotone{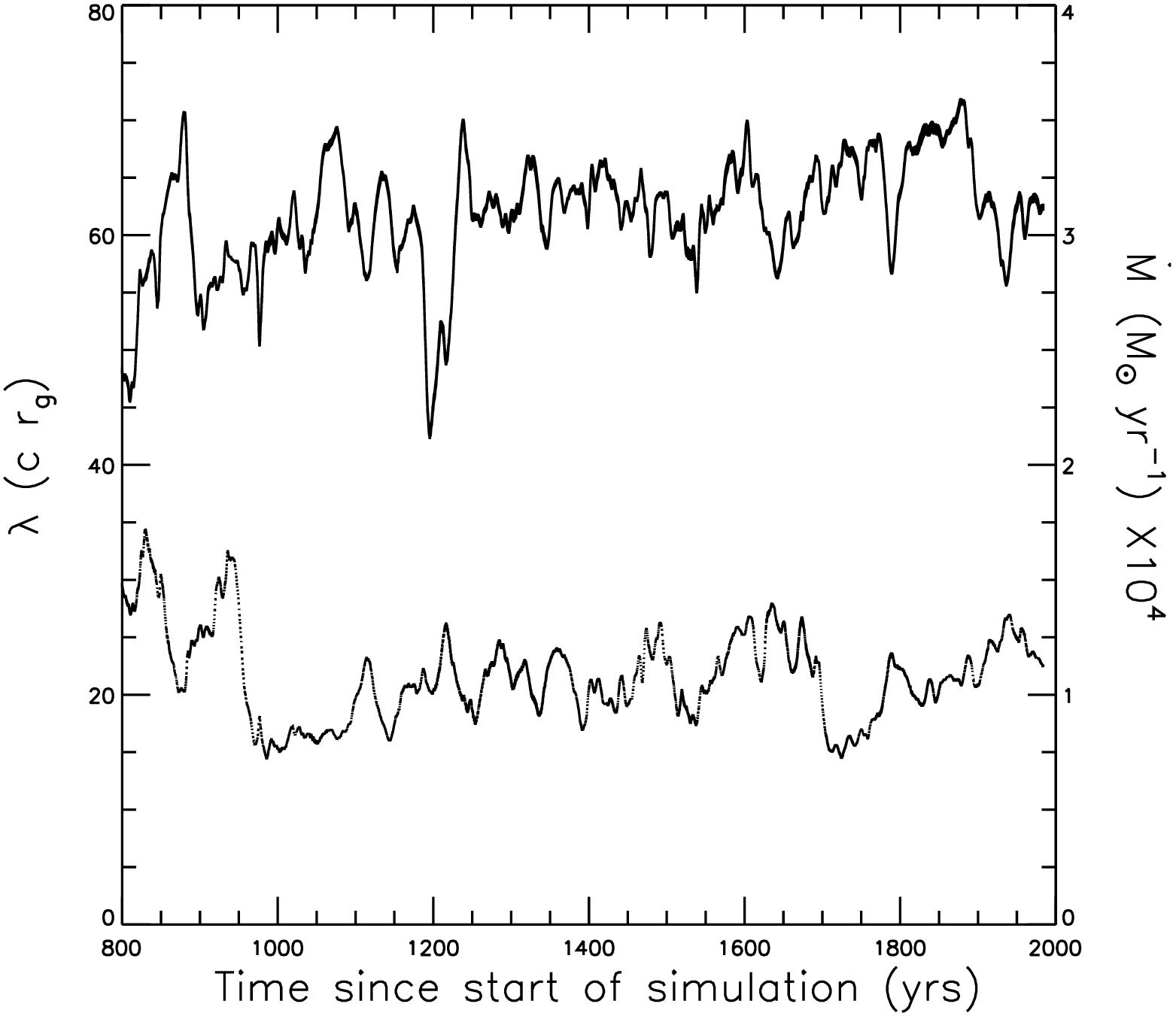}

\end{document}